\documentclass[12pt,a4paper]{article}

\usepackage{graphics}
\usepackage{latexsym}
\usepackage{amsmath}
\usepackage{amssymb}

\title{Like-sign dimuon charge asymmetry in the decay of 
$B\overline{B}$ pairs}
\author{B. Hoeneisen and C. Mar\'{\i}n}
\date{\small{Universidad San Francisco de Quito \\
	19 September 2000}}
\begin{document}
\maketitle

\begin{abstract}
\noindent
We discuss the CP-violating dimuon charge 
asymmetry of events 
$p \bar{p} \rightarrow B \bar{B} X \rightarrow \mu^\pm \mu^\pm X$
in the Standard Model.
Our conclusion is that the asymmetry is larger than 
previously expected and may reach a few percent for the
$\left( B_d^0, \bar{B}_d^0 \right)$ system. 
The analysis is 
also extended to the Two Higgs Doublet Model (Model II).
\end{abstract}


\section{Introduction}
We discuss the inclusive CP-violating dimuon charge asymmetry 
$A \equiv (N_{++} - N_{--})/(N_{++} + N_{--})$
in the Standard Model of events
$p \bar{p} \rightarrow B \bar{B} X \rightarrow \mu^\pm \mu^\pm X$
where the $B \bar{B}$ pair is either $B_d \bar{B}_d$ or
$B_s \bar{B}_s$. $N_{++}$ is the number of events
$p \bar{p} \rightarrow B \bar{B} X \rightarrow \mu^+ \mu^+ X$.
In particular we consider \textquotedblleft{indirect CP violation 
(or CP violation in the mixing)}"\cite{pdg} due to complex 
effects in $B \leftrightarrow \bar{B}$ mixing and decay.

\section{Indirect CP violation}
Let us review the standard formalism of indirect CP 
violation\cite{bh} but without making the usual 
approximations valid when CP violation is 
\textquotedblleft{small}". We take the hamiltonian in 
the $B_d^0 \leftrightarrow \bar{B}_d^0$
(or $B_s^0 \leftrightarrow \bar{B}_s^0$,
$D^0 \leftrightarrow \bar{D}^0$, or
$K^0 \leftrightarrow \bar{K}^0$) basis as
\begin{equation}
H \equiv M - \frac{i}{2} \Gamma \equiv
\left[ \begin{array}{cc} 
m & M_{12} \\
M_{12}^* & m 
\end{array}
\right] -
\frac{i}{2} 
\left[ \begin{array}{cc}
\Gamma & \Gamma_{12} \\
\Gamma_{12}^* & \Gamma 
\end{array}
\right],
\label{H}
\end{equation}
where the matrices $M$ and $\Gamma$ are hermitian. 
The hamiltonian $H$ itself is not hermitian since
the $B$ mesons do decay. The matrix elements of the
hamiltonian are obtained from second order perturbation
theory:
\begin{align}
M_{\alpha \beta} & =  m \delta_{\alpha \beta} +
\left< \alpha \mid H_{SW} \mid \beta \right> -
P \sum_{\xi \neq 1,2}{\frac{\left< \alpha \mid H_W \mid \xi \right>
\left< \xi \mid H_W \mid \beta \right>}{E_\xi - m}}
\label{M}, \\
\Gamma_{\alpha \beta} & =  2 \pi \sum_{\xi \neq 1,2}{
\left< \alpha \mid H_W \mid \xi \right>
\left< \xi \mid H_W \mid \beta \right> 
\delta (E_\xi - m)},
\label{Gamma}
\end{align}
where $H_W$ is the Standard Model weak interaction,
$H_{SW}$ is a superweak interaction (which we consider
no further), and $P$ denotes \textquotedblleft{principal part}".

The diagonal elements of $H$ are assumed equal due to
CPT invariance. The argument goes as follows: $M_{11}$ is the 
amplitude for a $B^0$ to remain a $B^0$. Applying CPT we obtain
the amplitude for a $\bar{B}^0$ to remain a $\bar{B}^0$, 
\textit{i.e.} $M_{22}$. So, if CPT invariance holds,
we obtain $M_{11} = M_{22} \equiv m$. 
Likewise, $\Gamma_{11}$ is the
probability per unit time for the decay $B^0 \rightarrow \sum{\xi}$.
Applying CPT we obtain the probability per unit time for
$\sum{\bar{\xi}} \rightarrow \bar{B}^0$. From Equation (\ref{Gamma})
for this process, and changing the order of the two brackets, we 
obtain the probability per unit time for 
$\bar{B}^0 \rightarrow \sum{\bar{\xi}}$, \textit{i.e.}
$\Gamma_{22}$. So, if CPT invariance holds, we obtain
$\Gamma_{11} = \Gamma_{22} \equiv \Gamma$.

The solution to the equation 
$i \partial \psi / \partial t = H \psi$ with
$\psi ^ T \equiv \left( B^0(t), \bar{B}^0(t) \right)$ is
\begin{eqnarray}
B^0(t) = \frac{1}{2} \left\{ s_+(t) + s_-(t) \right\} B^0(0)
+ \frac{1 - \varepsilon}{1 + \varepsilon} 
\cdot \frac{1}{2}
\left\{ s_+(t) - s_-(t) \right\}
\bar{B}^0(0), \nonumber \\
\bar{B}^0(t) = \frac{1 + \varepsilon}{1 - \varepsilon}
\cdot \frac{1}{2}
\left\{ s_+(t) - s_-(t) \right\} B^0(0) +
\frac{1}{2} \left\{ s_+(t) + s_-(t) \right\} \bar{B}^0(0),
\label{mixing}
\end{eqnarray}
where
\begin{eqnarray}
s_-(t)& = & \exp(-imt) \exp(-\Gamma t/2) 
\exp(i \Delta M t/2) \exp(\Delta \Gamma t/4), \nonumber \\
s_+(t) & = & \exp(-imt) \exp(-\Gamma t/2) 
\exp(-i \Delta M t/2) \exp(- \Delta \Gamma t/4),
\label{s}
\end{eqnarray}
\begin{equation}
\frac{1 - \varepsilon}{1 + \varepsilon} \equiv
\frac{\Delta M - \frac{i}{2} \Delta \Gamma}
{2 \left( M_{12}^* - \frac{i}{2} \Gamma_{12}^* \right)} =
\frac{2 \left( M_{12} - \frac{i}{2} \Gamma_{12} \right)}
{\Delta M - \frac{i}{2} \Delta \Gamma}.
\label{epsilon}
\end{equation}
The phase of $(1 - \varepsilon)/(1 + \varepsilon)$
is arbitrary: it can be changed by redefining the phase of 
$\bar{B}^0(0)$. Observables depend on the absolute value
of $(1 - \varepsilon)/(1 + \varepsilon)$, or equivalently
on
\begin{equation}
\alpha \equiv \frac{Re(\varepsilon)}
{1 + \left| \varepsilon \right|^2}.
\label{alpha}
\end{equation}
For the same reason, we can multiply 
$M_{12}$ and $\Gamma_{12}$ by a common phase-factor.
Only the relative phase is observable:
\begin{equation}
\angle \frac{\Gamma_{12}}{M_{12}} \equiv \varphi.
\label{phase}
\end{equation}
We introduce the notation
\begin{equation}
x \equiv \frac{\Delta M}{\Gamma}; 
y \equiv \frac{\Delta \Gamma}{2 \Gamma}.
\label{xy}
\end{equation}
Then the probability that a $\bar{B}^0$ decays as a
$B^0$ is
\begin{eqnarray}
\chi & = & \frac
{
\int_{0}^{\infty}{\left| \frac{1 - \varepsilon}
{1 + \varepsilon} \right|^2
\frac{1}{4} \left| s_+ - s_- \right|^2 dt}
}
{
\int_{0}^{\infty}{\left| \frac{1 - \varepsilon}
{1 + \varepsilon} \right|^2
\frac{1}{4} \left| s_+ - s_- \right|^2 dt }
+ \int_{0}^{\infty}{\frac{1}{4} \left| s_+ + s_- \right|^2 dt}
} \nonumber \\
& = & \frac{\left( x^2 + y^2 \right) 
\left( \frac{1}{2} - \alpha \right)}
{1 + x^2 + 2 \alpha \left( 1 - y^2 \right)}.
\label{chi} 
\end{eqnarray}
Similarly, the probability that a $B^0$ decays as a $\bar{B}^0$ is
\begin{equation}
\bar{\chi} = \frac{\left( x^2 + y^2 \right)
\left( \frac{1}{2} + \alpha \right)}
{1 + x^2 - 2 \alpha \left( 1 - y^2 \right) }.
\label{chibar}
\end{equation}
Finally, let us relate $M_{12}$ and $\Gamma_{12}$ with
$\Delta M$ and $\Delta \Gamma$:
\begin{equation}
\frac{\alpha}{1 + 4 \alpha^2} =
- \frac{Im \left\{ \Gamma_{12} / M_{12} \right\}}
{4 + \vert \Gamma_{12} / M_{12} \vert^2},
\label{alpha1}
\end{equation}
\begin{equation}
Im \left( \frac{\Gamma_{12}}{M_{12}} \right) =
- 4 \alpha \frac{1 + \left( \Delta \Gamma / (2 \Delta M) \right)^2}
{1 + \left( \alpha \Delta \Gamma / \Delta M \right)^2},
\label{Im}
\end{equation}
\begin{equation}
\tan ( \varphi ) = \frac{- \alpha}{1 - 4 \alpha^2} \cdot
\frac{4 \Delta M^2 + \Delta \Gamma^2}{\Delta M \Delta \Gamma},
\label{tan}
\end{equation}
\begin{equation}
\left| M_{12} \right|^2 = \frac{1}{4} \frac{1}{1 - 4 \alpha^2}
\Delta M^2 +
\frac{1}{4} \frac{\alpha^2}{1 - 4 \alpha^2} \Delta \Gamma^2,
\label{M12}
\end{equation}
\begin{equation}
\left| \Gamma_{12} \right|^2 = 
\frac{4 \alpha^2}{1 - 4 \alpha^2} \Delta M^2 + 
\frac{1}{4} \frac{1}{1 - 4 \alpha^2} \Delta \Gamma^2.
\label{Gamma12}
\end{equation}
These equations are exact for the 
$\left( B_d^0, \bar{B}_d^0 \right)$ and
$\left( B_s^0, \bar{B}_s^0 \right)$ systems separately.
As a cross check, for $\vert \alpha \vert \ll 1$
we recover the well known results
$\Delta M \approx 2 \vert M_{12} \vert$ and
$A \approx - 4 \alpha \approx Im \left( \Gamma_{12} / M_{12} \right)$.

\section{$\Gamma_{12}$ and $M_{12}$ in the Standard Model}
From the box diagrams\cite{pdg} of 
$B^0 \leftrightarrow \bar{B}^0$ mixing and Equation (\ref{M})
for the Standard Model, we obtain\cite{bh}
\begin{equation}
x_q \equiv \frac{\Delta M}{\Gamma} \approx
\frac{2 \vert M_{12} \vert}{\Gamma} \approx
\frac{\eta' f_B^2 G_F^2 m_W^2 m_B}
{8 \pi^2 \Gamma}
\vert V_{tq}^* \vert^2 \alpha \left( m_t^2 / m_W^2 \right)
\label{x}
\end{equation}
where
\begin{equation}
\alpha \left( x_t \right) \equiv
\frac{x_t^3 - 11 x_t^2 + 4 x_t}{4 \left( 1 - x_t \right)^2}
- \frac{3 x_t^3 \ln \left( x_t \right)}{2 \left( 1 - x_t \right)^3}.
\label{a}
\end{equation}

\begin{figure}
\begin{center}
\vspace*{-8.2cm}
\scalebox{0.6}
{\includegraphics[0in,0.5in][8in,9.5in]{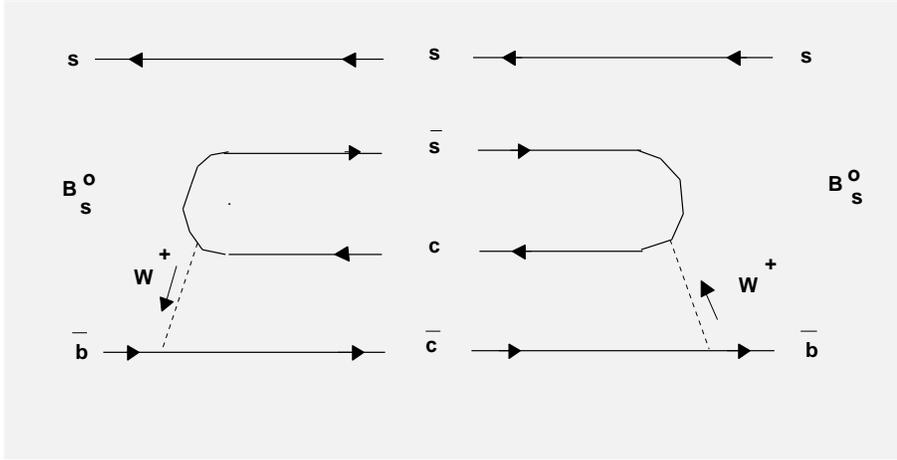}}
\vspace*{0.7cm}
\caption{A tree level Feynman diagram contributing to
the $B_s^0$ meson decay rate 
$\Gamma_{11} \equiv \Gamma \propto 
\left< B_s^0 \mid H_W \mid \xi \right> 
\left< \xi \mid H_W \mid B_s^0 \right>$. The intermediate
quarks can also be $\bar{d}$, $u$ or $\bar{u}$. There are
also lepton channels.}
\label{Gamma11.fig}
\end{center}
\end{figure}

\begin{figure}
\begin{center}
\vspace*{-8.2cm}
\scalebox{0.6}
{\includegraphics[0in,0.5in][8in,9.5in]{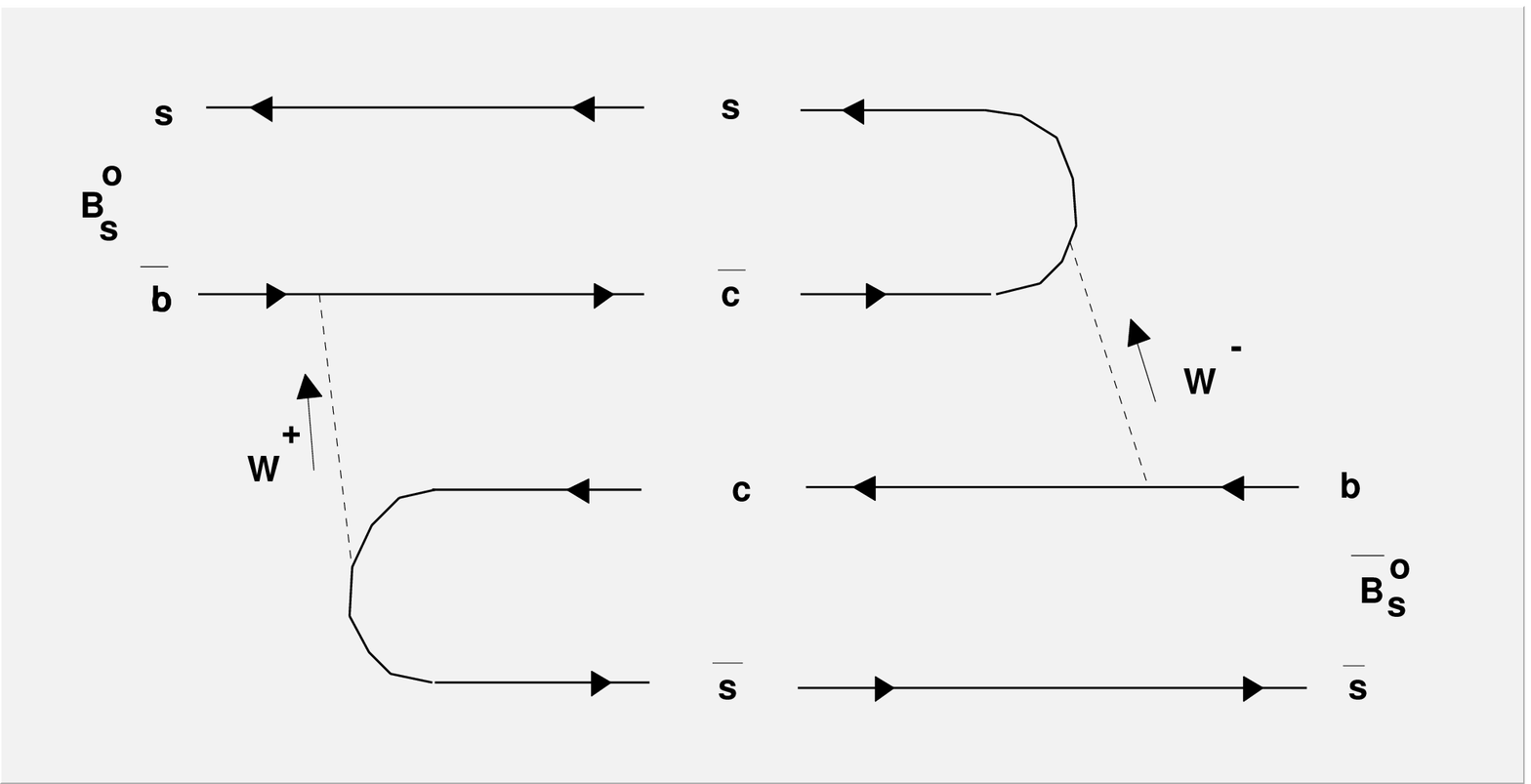}}
\vspace*{0.7cm}
\caption{A tree level Feynman diagram contributing to
$\Gamma_{12} \propto 
\left< B_s^0 \mid H_W \mid \xi \right>
\left< \xi \mid H_W \mid \bar{B}_s^0 \right>$.
The intermediate quarks can also be $\bar{u}$ or $u$.}
\label{Gamma12.fig}
\end{center}
\end{figure}

Let us now consider the absorptive matrix elements 
given by Equation (\ref{Gamma}). For $\alpha = \beta$ 
Equation (\ref{Gamma}) is \textquotedblleft{Fermi's
golden rule}". The tree level Feynman diagrams 
are shown in Figures \ref{Gamma11.fig} and \ref{Gamma12.fig}.

We arrive at \textbf{three} different predictions for the 
CP-violating asymmetries $A_d$ and $A_s$ in the Standard 
Model:

1. If we require that intermediate quark \textbf{spins 
and momenta} match in the diagrams of Figure \ref{Gamma12.fig}, 
then we obtain negligible asymmetries as discussed by 
J. Hagelin\cite{Hagelin}. This result would be correct if
quarks were observable.

2. However quarks are not observable. Hadrons are
observable. Due to hadronization, \textit{i.e.} due to gluons,
\textbf{the momenta need not match at the quark level} 
and we obtain the approximate results shown in Table \ref{table1}.
(All input numerical data for these calculations 
not specified in the Tables were obtained from 
\cite{pdg}.) A similar point was emphasized by T. Altomari,
L. Wolfenstein, and J.D. Bjorken\cite{Bjorken} so that their
\textquotedblleft{conclusion is that a reasonable
estimate of the asymmetry lies between $10^{-3}$ and $10^{-2}$
but that neither the sign nor the magnitude can be
reliably calculated}". That conclusion is in agreement
with Table \ref{table1}. 

3. If, due to gluon exchange in the hadronization process, 
it were a good approximation to \textbf{neglect both the
requirements of spin and momentum match}, then we obtain 
approximately the results shown in Table \ref{table2}. 
As an example, suppose that the c-quark
in the left hand diagram of Figure \ref{Gamma12.fig}
has \textquotedblleft{spin up}", and the c-quark in the right
hand diagram has 
\textquotedblleft{spin down}". This miss-match of spins of
the quarks need not imply a miss-match at the
hadron level, as can be seen, for example, in the
case in which both quarks hadronize into a scalar meson.

Which of the three predictions, if any, is correct?

\section{Discussion}
Consider diagrams of the form shown in Figure 
\ref{Gamma11.fig}. Requiring match of spins of the quarks
(and weighting hadronic modes by a factor 3 for color)
we obtain an inclusive branching fraction
$B(b \rightarrow \mu X) = 0.16$ to be compared with the
experimental value
$B(b \rightarrow \mu X) = 0.103 \pm 0.005$\cite{pdg}. 
This is the 
well known \textquotedblleft{baffling semi-leptonic
branching fraction of B mesons}"\cite{Bigi}. But quark 
spins are not observable. Only hadrons are observable.
So the hadronization process, \textit{i.e.} gluons,
enhance the decay rate $\Gamma_{11}$ by a factor
$\approx 2$ to account for the observed drop in 
semi-leptonic branching fraction.

\begin{figure}
\begin{center}
\vspace*{-8.2cm}
\scalebox{0.6}
{\includegraphics[0in,0.5in][8in,9.5in]{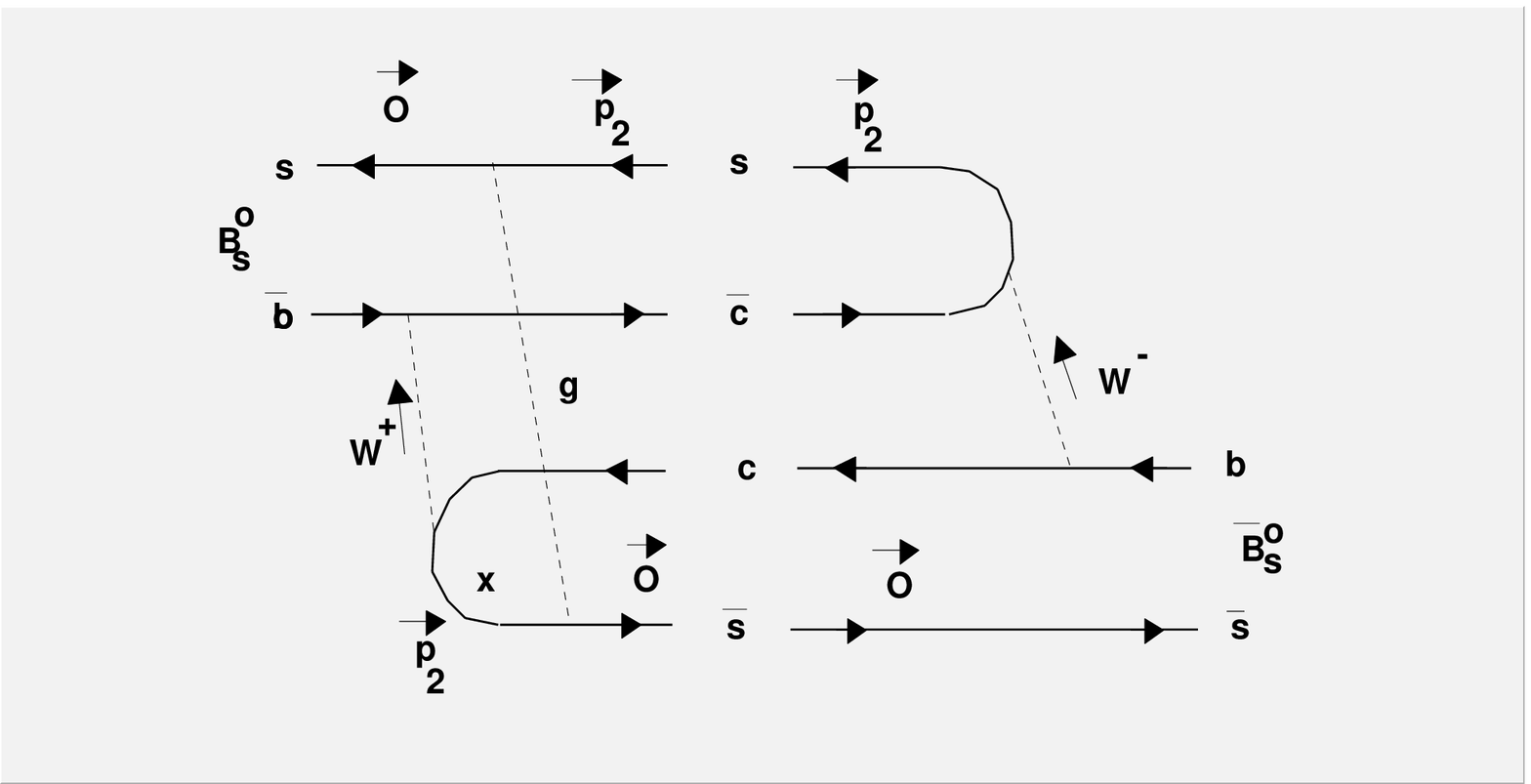}}
\vspace*{0.7cm}
\caption{Another Feynman diagram contributing to
$\Gamma_{12} \propto
\left< B_s^0 \mid H_W \mid \xi \right>
\left< \xi \mid H_W \mid \bar{B}_s^0 \right>$.
The intermediate quarks can also be $\bar{u}$ or $u$.}
\label{gluon.fig}
\end{center}
\end{figure}

Let us now consider the momentum miss-match of the 
spectator quarks in the diagrams of Figure \ref{Gamma12.fig}.
To obtain matching of momenta we need at least one gluon as
shown in Figure \ref{gluon.fig}.
The amplitude of the diagram of Figure \ref{gluon.fig}
is \textquotedblleft{enhanced}" 
with respect to the amplitude of the
diagram of Figure \ref{Gamma12.fig} by a factor of order
\begin{equation}
\frac{\Gamma_{12}(3)}{\Gamma_{12}(2)} \approx
\frac{A_{x}(3)}{A_{x}(2)} \approx
\frac{\alpha_s m_B f_B^2}{8} \cdot P_x \cdot P_g.
\label{enhancement}
\end{equation}
The numbers in parenthesis refer to the Figures, and
$x \equiv s$ or $d$ for the $B_s^0$ or $B_d^0$ systems
respectively. 
The factor $m_B f_B^2$ was introduced (somewhat arbitrarily)
on dimensional grounds.
The gluon propagator is 
$P_g \approx 1/(2 m_x \vert \vec{p}_2 \vert )$. 
We take $\vert \vec{p}_2 \vert$ to be of order $m_B/3$
(see definition of $\vec{p}_2$ and $x$ in Figure \ref{gluon.fig}).
If the outgoing quarks were free, the quark propagator $P_x$ 
would be on-shell and would diverge. Since the 
outgoing quarks are confined to dimensions of order $1$ Fermi, 
we replace the quark propagator by $P_x \approx 1/m_\pi$.
Then the order of magnitude estimate of the 
\textquotedblleft{enhancement factor}" (\ref{enhancement}) is
\begin{equation}
\frac{\Gamma_{12}(3)}{\Gamma_{12}(2)} \approx
\frac{A_{x}(3)}{A_{x}(2)} \approx
\frac{3 \alpha_s f_B^2}{16 m_\pi m_x}
\label{approx_enhancement}
\end{equation}
which is $\approx 0.08$ for the $B_s^0$ system,
and $\approx 1.6$ for the $B_d^0$ system.

So, which of the three predictions, if any,
is correct? Prediction 1 by Hagelin includes a 
restriction on $\vert \vec{p}_2 \vert$ due to a miss-match 
of momenta of the spectator quark, and also requires
matching spins. We have seen that this momentum miss-match
can be \textquotedblleft{fixed}" by gluons (at no cost for
the $B_d^0$ system). Therefore we arrive at Prediction
2 given in Table \ref{table1} which does not require momenta
to match but still requires matching spins. We have seen 
that hadronization relaxes the need to match spins
(as required by the semi-leptonic branching fraction
of $B$ mesons). Then
we arrive at Prediction 3 given in Table \ref{table2}.
Since $x_d = 0.723 \pm 0.032$\cite{pdg} we obtain from
Table \ref{table2}, $A_d \approx 1\%$ to $3\%$ for the
allowed range of $V_{ub}$\cite{pdg}.
But for the $B_d^0$ system the gluon in the diagram
of Figure \ref{gluon.fig} can even enhance the asymmetry
\textbf{above} the prediction given in Table \ref{table2}.
Note that relaxing the requirement of matching spins we obtain
a large $\Delta \Gamma_s / \Gamma$ as predicted in
\cite{Browder}, leading to interesting experimental 
consequences\cite{Dunietz}.

\begin{table}
\begin{center}
\begin{tabular}{|l|rrrrrrr|}\hline
$\gamma$ & $22.5^0$ & $45^0$ & $67.5^0$ & $90^0$ &
$112.5^0$ & $135^0$ & $157.5^0$ \\
\hline
$\beta$ & $12.9^0$ & $20.5^0$ & $22.7^0$ & $21.1^0$ & 
$17.2^0$ & $12.1^0$ & $6.2^0$ \\
$\varphi_d$ & $-41^0$ & $-78^0$ & $-106^0$ & $-127^0$ & $216^0$ &
$203^0$ & $191^0$ \\
$A_d$ & $-.18\%$ & $-.33\%$ & $-.36\%$ &
$-.32\%$ & $-.25\%$ & $-.17\%$ & $-.08\%$ \\
$A_s$ & $.04\%$ & $.04\%$ &
$.05\%$ & $.05\%$ & $.04\%$ & $.04\%$ & $.03\%$ \\
$x_d$ & $0.48$ & $0.66$ & $0.93$ & $1.26$ & $1.58$ &
$1.85$ & $2.04$ \\
$x_s$ & $22$ & $22$ & $22$ & $22$ & 
$22$ & $22$ & $22$ \\
$\Delta \Gamma_d / \Gamma$ &  $0.1\%$ & $0.0\%$ & $0.1\%$ &
 $0.3\%$ & $0.5\%$ & $0.7\%$ & $0.9\%$ \\
$\Delta \Gamma_s / \Gamma$ & $3.3\%$ & $3.2\%$ &
 $3.0\%$ & $2.8\%$ & $2.7\%$ & $2.5\%$ & $2.5\%$ \\
\hline
\end{tabular}
\end{center}
\caption{Standard Model calculation of the dimuon
charge asymmetries $A_d$ and $A_s$, and other 
parameters, as a function of the angles $\gamma$ and
$\beta$ of the unitarity triangle\cite{pdg}, for
$V_{ub} = 0.0036$, $\eta' = 1.09$, and $f_B = 0.18$GeV.
The following parameters define the Cabbibo-Kobayashi-Maskawa
(CKM) matrix in the Wolfenstein parametrization\cite{pdg}:
$\lambda = 0.2235$, $A = 0.8357$, 
$\left[ \rho^2 + \eta^2 \right]^{1/2} = 0.3854$ and
$\gamma \equiv \arctan(\eta / \rho)$ listed above.
Momenta of the s-quarks in the diagram of Figure 
\ref{Gamma12.fig} \textbf{are not} required to match.
Spins \textbf{are} required to match. $m_b = 4.25$GeV/c$^2$, 
$m_c = 1.25$GeV/c$^2$,
$m_u = 0.00325$GeV/c$^2$, $m_d = 0.006$GeV/c$^2$, 
and $m_s = 0.115$GeV/c$^2$.}
\label{table1}
\end{table}

\section{The Two Higgs Doublet Model}
We have repeated the calculations using the 
Two Higgs Doublet Model (Model II)\cite{Higgs}.
The results are shown in Table \ref{table3}
(which requires matching intermediate quark spins
but not momenta).
Comparing with Table \ref{table1} we note that
the predictions for the asymmetry $A_d$ are not
significantly changed by this extension of the
Standard Model.

\section{Conclusions}
Our conclusion is that the inclusive CP-violating 
like-sign 
dimuon charge asymmetry in the Standard Model is 
larger than previously expected\cite{Bjorken}.
For the $(B_d^0, \bar{B}_d^0)$ system
the asymmetry is
probably between $\approx 1\%$ and $3\%$, and
may even be enhanced by gluons above this value.
Such a large asymmetry would be within the reach of the
next run of the D$\emptyset$ and CDF experiments at the
FERMILAB Tevatron collider. 

\begin{table}
\begin{center}
\begin{tabular}{|l|rrrrrrr|}\hline
$\gamma$ & $22.5^0$ & $45^0$ & $67.5^0$ & $90^0$ &
$112.5^0$ & $135^0$ & $157.5^0$ \\
\hline
$\beta$ & $12.9^0$ & $20.5^0$ & $22.7^0$ & $21.1^0$ & 
$17.2^0$ & $12.1^0$ & $6.2^0$ \\
$\varphi_d$ & $102^0$ & $67^0$ & $47^0$ & $33^0$ & $23^0$ &
$14^0$ & $7^0$ \\
$A_d$ & $1.3\%$ & $1.9\%$ & $1.8\%$ &
$1.5\%$ & $1.1\%$ & $0.7\%$ & $0.4\%$ \\
$A_s$ & $-.06\%$ & $-.11\%$ &
$-.14\%$ & $-.15\%$ & $-.14\%$ & $-.10\%$ & $-.06\%$ \\
$x_d$ & $0.48$ & $0.66$ & $0.93$ & $1.26$ & $1.58$ &
$1.85$ & $2.04$ \\
$x_s$ & $22$ & $22$ & $22$ & $22$ & 
$22$ & $22$ & $22$ \\
$\Delta \Gamma_d / \Gamma$ &  $0.1\%$ & $0.5\%$ & $1.6\%$ &
 $2.9\%$ & $4.1\%$ & $5.2\%$ & $6.0\%$ \\
$\Delta \Gamma_s / \Gamma$ & $39\%$ & $39\%$ &
 $38\%$ & $36\%$ & $35\%$ & $34\%$ & $33\%$ \\
\hline
$\beta$ & $22.4^0$ & $31.7^0$ & $32.2^0$ & $28.4^0$ & 
$22.5^0$ & $15.4^0$ & $7.8^0$ \\
$A_d$ & $2.6\%$ & $3.0\%$ & $2.5\%$ &
$1.9\%$ & $1.3\%$ & $0.8\%$ & $0.4\%$ \\
$A_s$ & $-0.1\%$ & $-0.2\%$ &
$-0.2\%$ & $-0.2\%$ & $-0.2\%$ & $-0.1\%$ & $-0.1\%$ \\
$x_d$ & $0.32$ & $0.58$ & $0.96$ & $1.41$ & $1.86$ &
$2.25$ & $2.50$ \\
$x_s$ & $22$ & $22$ & $22$ & $22$ & 
$22$ & $22$ & $22$ \\
$\Delta \Gamma_d / \Gamma$ &  $0.4\%$ & $0.5\%$ & $1.9\%$ &
 $3.7\%$ & $5.5\%$ & $7.1\%$ & $8.1\%$ \\
$\Delta \Gamma_s / \Gamma$ & $40\%$ & $39\%$ &
 $38\%$ & $36\%$ & $34\%$ & $33\%$ & $32\%$ \\
\hline
\end{tabular}
\end{center}
\caption{Same as Table \ref{table1} except that the spins
(in addition to the momenta) of the intermediate quarks
in Figure \ref{Gamma12.fig} are \textbf{not} required to match.
For the first half of the Table 
$\left[ \rho^2 + \eta^2 \right]^{1/2} = 0.3854$ and
$V_{ub} = 0.0036$. For the second half of the Table
$\left[ \rho^2 + \eta^2 \right]^{1/2} = 0.54$ and 
$V_{ub} = 0.0050$. At $\gamma = 0^0$ and $180^0$,
$A_d = 0$ and $A_s = 0$.}
\label{table2}
\end{table}

\begin{table}
\begin{center}
\begin{tabular}{|l|rrrrrrr|}\hline
$\gamma$ & $22.5^0$ & $45^0$ & $67.5^0$ & $90^0$ &
$112.5^0$ & $135^0$ & $157.5^0$ \\
\hline
$\beta$ & $12.9^0$ & $20.5^0$ & $22.7^0$ & $21.1^0$ & 
$17.2^0$ & $12.1^0$ & $6.2^0$ \\
$\varphi_d$ & $-43^0$ & $-80^0$ & $-109^0$ & $-130^0$ & $248^0$ &
$200^0$ & $187^0$ \\
$A_d$ & $-.17\%$ & $-.31\%$ & $-.33\%$ &
$-.29\%$ & $-.22\%$ & $-.14\%$ & $-.05\%$ \\
$A_s$ & $.28\%$ & $.27\%$ &
$.26\%$ & $.25\%$ & $.24\%$ & $.23\%$ & $.23\%$ \\
$x_d$ & $0.48$ & $0.66$ & $0.93$ & $1.26$ & $1.58$ &
$1.85$ & $2.04$ \\
$x_s$ & $22$ & $22$ & $22$ & $22$ & 
$22$ & $22$ & $22$ \\
$\Delta \Gamma_d / \Gamma$ &  $0.1\%$ & $0.0\%$ & $0.1\%$ &
 $0.3\%$ & $0.5\%$ & $0.7\%$ & $0.8\%$ \\
$\Delta \Gamma_s / \Gamma$ & $2.3\%$ & $2.4\%$ &
 $3.0\%$ & $2.6\%$ & $2.5\%$ & $2.3\%$ & $2.1\%$ \\
\hline
\end{tabular}
\end{center}
\caption{Same as Table \ref{table1} but for the Two Higgs
Doublet Model (Model II) with $m_H^\pm = 45$GeV/c$^2$ and
$\tan \beta = 30$, instead of the Standard Model.}
\label{table3}
\end{table}

\end{document}